\documentclass[aps,prd,twocolumn,nofootinbib,showpacs,superscriptaddress,nofootinbib]{revtex4-1}

\usepackage{bm}
\usepackage{hyperref}
\usepackage{framed}
\usepackage{subfigure}
\usepackage{graphicx}
\usepackage{epstopdf}
\usepackage{amssymb}
\usepackage{amsmath}
\usepackage{color}
\def\be{\begin{equation}}
\def\ee{\end{equation}}
\def\ba{\begin{align}}
\def\ea{\end{align}}

\def\mbf{\mathbf}
\def\tbf{\textbf}

\newcommand{\one}{\leavevmode\hbox{\small1\kern-3.8pt\normalsize1}}

\begin{document}

\title{Are there really `black holes' in the Atlantic Ocean?}

\author{Angus Prain}
\affiliation{Physics Department and STAR Research Cluster, Bishop's University, 2600 College St.,
Sherbrooke, Qu\'ebec, Canada J1M 1Z7}
\author{Valerio Faraoni}
\affiliation{Physics Department and STAR Research Cluster, Bishop's University, 2600 College St.,
Sherbrooke, Qu\'ebec, Canada J1M 1Z7}

\begin{abstract}

In this letter we point out some interpretational 
difficulties associated with concepts from General Relativity in a recent article which appeared in 
 \textit{J. Fluid Mech.}  \tbf{731} (2013) R4 where a 
 Lorentzian metric was 
defined for turbulent fluid flow and interpreted as being analogous to a black hole metric. We show that the  
similarity with black hole geometry is superficial at best while clarifying the nature of the black hole geometry and the work above with some examples.

\end{abstract}

\maketitle

\section{Introduction}

In a recent article `Coherent Lagrangian vortices: the black holes of 
turbulence' \cite{Haller:2013wpa} the authors 
 draw parallels between 
certain fluid dynamical configurations and certain 
properties of black hole geometry. The goal the article \cite{Haller:2013wpa} is to characterize 
vortex structures which remain coherent over long times. Some 
such vortices (known as the Agulhas rings) are observed in the 
South Atlantic and are believed to be relevant for the long range 
transport of water with relatively high salinity and temperature, and possibly 
also as moving oases for the food chain (\cite{Haller:2013wpa} and 
references therein).

Specifically, the authors of \cite{Haller:2013wpa}  describe how to 
associate a 1+1 dimensional  
\textit{Lorentzian} effective metric tensor to 
spatial (constant  time) snapshots of a fluid flow.  One  claim 
made in \cite{Haller:2013wpa} is that certain closed spatial 
curves (at fixed time) in the fluid flow are analogous to the 
`photon spheres'\footnote{The circular 
photon orbit around a black hole, which defines the photon 
sphere, is a space-like circular curve along which a photon 
(or any massless particle) can travel  
under the influence of gravity alone.} that 
exist around black holes.\footnote{The so-called 
`Schwarzschild black hole' is the unique static spherically 
symmetric vacuum solution to Einstein's gravitational 
field equations and is known  to describe the geometry 
outside of a spherically symmetric gravitating 
material body, including a  non-rotating black hole 
\cite{Carroll, Wald}.} 
A photon sphere necessarily occurs in a black hole space-time and 
here we draw intuition (for a subsequent two dimensional treatment) from the simplest black 
hole, which is described (in four dimensions) by the Schwarzschild 
metric (see below).   

While we we do not take issue with the interesting fluid-dynamical 
subject matter of \cite{Haller:2013wpa}, we wish to point out a 
number of conceptual difficulties associated with the geometric 
interpretation of the results and hopefully elucidate  some of 
the (perhaps counter-intuitive) features of Lorentzian geometry and 
in particular of black hole geometries. Specifically, we make the 
following clarifications: 
\begin{itemize} 
 \item The circular photon orbit (associated with the photon 
sphere) around a Schwarzschild black hole is 
\textit{not} a closed null geodesic. 

\item Closed null curves in General Relativity are extremely 
pathological and probably forbidden by reasonable physics arguments 
(for example, globally hyperbolic spacetimes do not admit closed 
null curves).
 
\item The existence of a photon sphere is not a necessary nor is it 
a  sufficient condition for the existence of a black hole.

\item The  circular photon orbit is \textit{not} circular - its 
3-dimensional  \textit{projection onto the fixed time slices of the 
Schwarzschild  geometry in spherical polar coordinates} is.
\item A singularity in the metric does not imply a singularity of 
the  geometry. A singularity of a metric coefficient in one coordinate system is a weak condition and does not indicate the presence of a real physical singularity. 
\end{itemize} 
Below we will expand on these points.

In Sec.~\ref{S:2} we explore the construction of the `fluid-metric' 
as defined in \cite{Haller:2013wpa} and its geometry. In 
Sec.~\ref{S:3} we review some relevant aspects of the  
Schwarzschild geometry, in particular the circular photon orbit 
and the photon sphere, while in Sec.~\ref{S:4} we present some 
examples of fluid flows which give rise to interesting Lorentzian 
geometries and which serve to illustrate our observations. 

\section{Describing fluids with Lorentzian geometry} \label{S:2}

In this section we briefly review the main theoretical result of 
Ref.~\cite{Haller:2013wpa} by deriving the $1+1$-dimensional 
Lorentzian metric associated with certain fluid configurations.

The primary mathematical result of the article 
\cite{Haller:2013wpa} is the identification and characterisation of 
certain closed curves along which the fluid flow is `coherent': 
closed curves of material flow which remain closely associated. 
Such curves are shown in \cite{Haller:2013wpa} to satisfy a 
differential equation which is interpreted as an integrability 
condition of a vector field which is null with respect to an 
auxiliary metric tensor. That is, one discovers a null vector field and subsequently integrates it to yield the integral curves which are null curves. We shall introduce some concepts from 
continuum mechanics to this end.

Consider a sufficiently smooth fluid flow  
velocity $\mbf{v}(t,\mbf{x})$ (for example, $\mbf{v}$ could be a 
solution of the Navier-Stokes equations, but this is not required). 
Then a fluid element propagates in time along the flow lines with a 
world line $\mbf{x}(t)$ which satisfies
\be
\mbf{\dot{x}}=\mbf{v} \left(t, \mbf{x} \right). \label{E:ODE}
\ee

The flow given by \eqref{E:ODE} represents a map $\phi_t$ from 
$\mathbb{R}^2$ to $\mathbb{R}^2$ (we 
restrict attention to a 2-dimensional fluid flow) 
where the reference configuration given by the initial condition $\mbf{x}(0)=\mbf{x}_0$ gets 
mapped to the solution to \eqref{E:ODE} at time $t$ 
\be
\phi_t:\mbf{x}_0 \longrightarrow \mbf{X}(t).
\ee
Here we have denoted the coordinates in the plane at time $t$  with a capital letter $\mbf{X}$ for clarity 
while reserving the lower case variable names for the 
initial un-deformed coordinates $\mbf{x}$.  In general, the 
flow $\phi_t$ will deform the lines of constant initial 
coordinate label, say $x=$const. or $y=$const., where 
$\mbf{x}=(x,y)$ are Cartesian coordinates, so that lines 
of constant initial condition $x_0$ or $y_0$ will, at some fixed time $t$, be deformed to some curved lines in the plane. We may characterise such a deformation with the so-called 
`deformation gradient' mixed two-point tensor $\mbf{F}$ 
whose components with respect to an initial and final 
coordinate basis are given by
\be
F^i {}_j:=\frac{\partial X^i(t)}{\partial x_0^j} \,.
\ee
Note the mixed-coordinate definition (two-point tensor 
structure) by observing that these components are given 
with respect to the two coordinate bases, initial and 
final in a specific way. We have, in 
coordinate-independent notation, 
\be
\mbf{F}= F^i{}_j \; \left( \frac{\partial}{\partial X^i} \otimes dx^j\right).
\ee
For example, and for later reference, in a polar 
coordinate basis we have
\be
F^i{}_j=\begin{pmatrix} \frac{\partial R(t)}{\partial r_0} 
& \frac{\partial R(t)}{\partial \theta_0} \\ 
\vspace{-2mm} \\
\frac{\partial \Theta(t)}{\partial r_0} & \frac{\partial 
\Theta(t)}{\partial \theta_0} \end{pmatrix} \,,
\ee
which is written in the more convenient orthonormal polar basis $e_r:=\partial_r$ and $e_\theta:=\partial_\theta /r$ 
($e_R:=\partial_R$ and $e_\Theta:=\partial_\Theta/R$) as
\be
F^a{}_b=\begin{pmatrix} \frac{\partial R(t)}{\partial r_0} 
& \frac{1}{r_0}\frac{\partial R(t)}{\partial \theta_0} \\ 
\vspace{-2mm} \\ R(t)\frac{\partial \Theta(t)}{\partial 
r_0} & \frac{R(t)}{r_0}\frac{\partial \Theta(t)}{\partial 
\theta_0} \end{pmatrix}.
\ee
The orthonormal basis\footnote{Here and below we have indicated the orthonormal basis with Latin indices from the beginning of the alphabet.} is preferable to the polar coordinate basis due to the possibility to raise and lower indices with the identity matrix, which coincides with the matrix of metric coefficients in such a basis. 

The `Right Cauchy-Green' deformation tensor $\mbf{C}$ 
carries the same deformation information modulo 
transformations which merely shift or rotate an initial 
configuration without genuine deformation and is defined 
as
\be
C^i{}_j:=F_k{}^i F^k{}_j=G_{km} \, g^{in}\frac{\partial 
X^m(t)}{\partial x_0^n}\frac{\partial X^k(t)}{\partial 
x_0^j} \,,
\ee
where $g_{ij}$ and  $G_{ij}$ are the metric coefficients in 
the two coordinate bases, respectively. In orthonormal 
bases we have
\be
C_{ab}=\sum_c \frac{\partial X^c(t)}{\partial x_0^a} 
\frac{\partial X^c(t)}{\partial x_0^b} \,.
\ee

We will be concerned with the eigenvectors $\xi_i$ 
and the eigenvalues $\lambda_i$ of $\mbf{C}$. Note 
that  $\mbf{C}$ as well as the eigenvectors and eigenvalues 
depend on the initial point $\mbf{x}_0$ as well as on the 
time $t$ at which we calculate the strains. Being the 
`square' of another tensor $\mbf{F}$ it is simple to show 
that the eigenvalues $\lambda_i(\mbf{x}_0,t)$ of $\mbf{C}$ 
satisfy
\be
0\leq\lambda_1(\mbf{x}_0,t)\leq  \lambda_2(\mbf{x}_0,t)
 \quad \text{for all}\; \mbf{x}_0,\; t
\ee
and that $\{\xi_i\}$ form an orthonormal basis for $\mathbb{R}^2$. 

Let $\lambda>0$ and define the so-called `generalised 
Green-Lagrange tensor' $\mbf{E}_\lambda$ in an orthonormal 
frame by
\be
E_{ab}:=\frac{1}{2}\left[C_{ab}-\lambda \delta_{ab}\right]. \label{E:E}
\ee
This object can be thought of as acting as a time-dependent 
bilinear form on the initial conditions space and depends on the 
constant $\lambda$. In the `moving orthonormal frame' defined by 
the eigenvectors of $\mbf{C}$ we can write
\be
E_{ab}=\frac{1}{2}\begin{pmatrix} \lambda_1-\lambda & 0 \\ 0 & \lambda_2-\lambda \end{pmatrix} \,. \label{E:Eeigen}
\ee

The authors of \cite{Haller:2013wpa} define an effective metric tensor induced by the action of $\mbf{E}$ 
in \eqref{E:E} on the space 
of initial conditions which, remarkably, can have Lorentzian 
signature (although the tensor $\mbf{E}$ is of the wrong type to be a metric, we gloss over this and talk about $\mbf{E}$ as `being' the metric tensor). In principle, this fact allows one to  draw parallels 
between this structure and a black hole metric with an associated 
photon sphere. 

\subsection{The meaning and geometry of the metric $E_{ab}$}

The first thing to notice about the tensor $\mbf{E}$ is 
that its signature can change from point to point and over 
time depending on the relative magnitude of the eigenvalues 
$\lambda_i(\mbf{x}_0,t)$ and is only Lorentzian at 
time 
$t$ if $\lambda_1(\mbf{x}_0,t)<\lambda<\lambda_2(\mbf{x}_0,t)$ for 
all points 
$\mbf{x}_0$. Assuming this `Lorentzian 
condition' on the eigenvalues at some time $t$, we can 
construct the two independent `E-null' vector fields 
$\eta_\pm$ in the orthonormal eigen-basis of $\mbf{C}$ as \cite{Haller:2013wpa}
\begin{align}
\eta_\pm &:=\frac{1}{\sqrt{\lambda_2-\lambda_1}} \left( 
\sqrt{\lambda_2-\lambda} ,\;\; \pm\sqrt{\lambda-\lambda_1}  \right)\\
&=\sqrt{\frac{\lambda_2-\lambda}{\lambda_2 
-\lambda_1}}\;\xi_1\pm\sqrt{\frac{\lambda 
-\lambda_1}{\lambda_2-\lambda_{  1}  }}\;\xi_2 \,,
\end{align}
where we have (arbitrarily but without loss of generality) 
normalised the vector fields to have unit 
norm in the background Euclidean metric. The integral 
curves of these vector fields are null curves with respect 
to the metric $\mbf{E}$. It is shown in \cite{Haller:2013wpa} that these null curves are curves for 
which the flow up to time $t$ \textit{uniformly} scales 
the tangent vectors to the curve.  That is, given an 
$\mbf{E}$-null curve $\gamma(s)$ which gets mapped under 
the flow $\phi_t$ to the curve $\phi_t(\gamma)(s)$, we have
\be
||\phi_t(\gamma)'(s)||^2=\lambda ||\gamma'(s)||^2
\ee
uniformly for all $s$ parametrizing the curve and where the prime 
denotes taking the tangent vector at the given point on the curve. 
Such closed curves which possess the uniform stretching property 
are intuitively `invariant curves' of the flow $\phi_t$; their 
fluid dynamical relevance is discussed in \cite{Haller:2013wpa} 
where the outermost curve in a family of such curves is said to define the boundary of a `coherent material vorticex' in 
2-dimensional flow.

\section{Photon spheres and black holes - what they are and what 
they are not} \label{S:3}

The authors of~\cite{Haller:2013wpa} refer to a photon sphere in an 
analogy with vortices through the Lorentzian metric $\mbf{E}$.  
A photon sphere is associated with a black hole spacetime metric.  
In General Relativity there exists a vast array of black 
hole and black hole-like solutions with varying degrees of 
physical sensibility. In the vacuum, spherically symmetric 
case in $3+1$-dimensions, there is a uniqueness theorem (Birkoff's 
theorem  \cite{Carroll, Wald}) which points us to the static 
Schwarzschild  black hole metric
\be\label{Schwarzschild}
ds^2=-\left(1-\frac{2GM}{ r}\right)dt^2 
+\left(1-\frac{2GM}{r}\right)^{-1}dr^2+r^2 d\Omega_{(2)}^2
\ee
here written in standard spherical polar coordinates,  where 
$d\Omega_{(2)}^2=d\theta^2 +\sin^2 \theta \, d\varphi^2$ is the 
metric on the unit 2-sphere, $G$ is Newton's constant, $M$ is the 
black hole mass, and units  in which the speed of light is unity 
are used, as is common in relativity. The metric 
(\ref{Schwarzschild}) describes the simplest black hole spacetime 
in $3+1$ dimensions. A crucial feature of this 
metric is the existence of an horizon -- a lower-dimensional region 
on which the metric is degenerate and at which time and space `swap 
their roles'. Without getting into technical details, a precise 
statement is that the norm of the timelike Killing vector field 
associated with the time symmetry changes sign, becoming a 
space-like vector field at a co-dimension~2 hyper-surface generated 
by null geodesics of the metric~(\ref{Schwarzschild}) and known as 
the {\em event horizon}.  The event horizon is the key 
structure which,  if present in a gravitational configuration, 
indicates the presence of a black hole. In the above Schwarzschild 
case, the time-like Killing vector is simply the time direction 
$K_t:=\partial_t$, with norm
\be
||K_t||^2=-\left(1-\frac{2GM}{r}\right)
\ee
which changes sign at $r_s=2GM$, a value of the radial 
coordinate known as the {\em Schwarzschild radius} or black hole 
radius \cite{Carroll, Wald}. This is the radius of the event 
horizon. Without an event horizon, a 
spacetime cannot be sensibly said to contain a black 
hole.\footnote{A slight {\em caveat} to this classification is 
that it might be possible for black hole horizons to 
evaporate through a semi-classical process known as 
Hawking evaporation. In this case the horizon is not a true 
event horizon since it lives for only a finite amount of 
time.}

\subsection{The `circular' null geodesic of the Schwarzschild black hole}

In a careful relativistic treatment \cite{Carroll, 
Wald}, one can show that the effective potential for 
`free' (geodesic) radial motion in the Schwarzschild 
geometry is given by the effective 
1-dimensional potential
\be\label{effectivepotential}
V(r)=-\epsilon 
\, \frac{GM}{r}+\frac{L^2}{2r^2}-\frac{GML^2}{r^3} \,,
\ee
where $\epsilon=+1$ for massive particles and $\epsilon=0$ 
for massless particles,  
and $L$ is the angular momentum\footnote{For massive 
particles $L$ represents the angular momentum per unit 
mass.} of the particle. The first two terms  on the 
right hand side of eq.~(\ref{effectivepotential}) are identical 
to the two terms which appear in the effective potential of the Newtonian treatment \cite{Goldstein}, 
while the third one is a 
correction due to General Relativity. Circular orbits exist 
at radii  $r_\text{orbit}$ for which $V'(r_\text{orbit})=0$ 
and we have
\be
r_\text{orbit}=\left(L\pm\sqrt{L^2-12\, \epsilon 
\, G^2M^2}\right)\frac{L}{2\epsilon \, GM} \,.
\ee
In the massless  limit $\epsilon=0$, there exists a
single  orbit at $3GM$, which is unstable, while in the  
massive case there exist two orbits, one stable and the 
other (at smaller radius) unstable.  In 
contrast, in Newtonian gravity (in which the term 
proportional to $r^{-3}$ is absent in the effective 
potential $V(r)$), there exists a single stable circular orbit for 
massive particles and no circular orbit for massless 
particles.  

\begin{figure}
\centering
\includegraphics[scale=0.3]{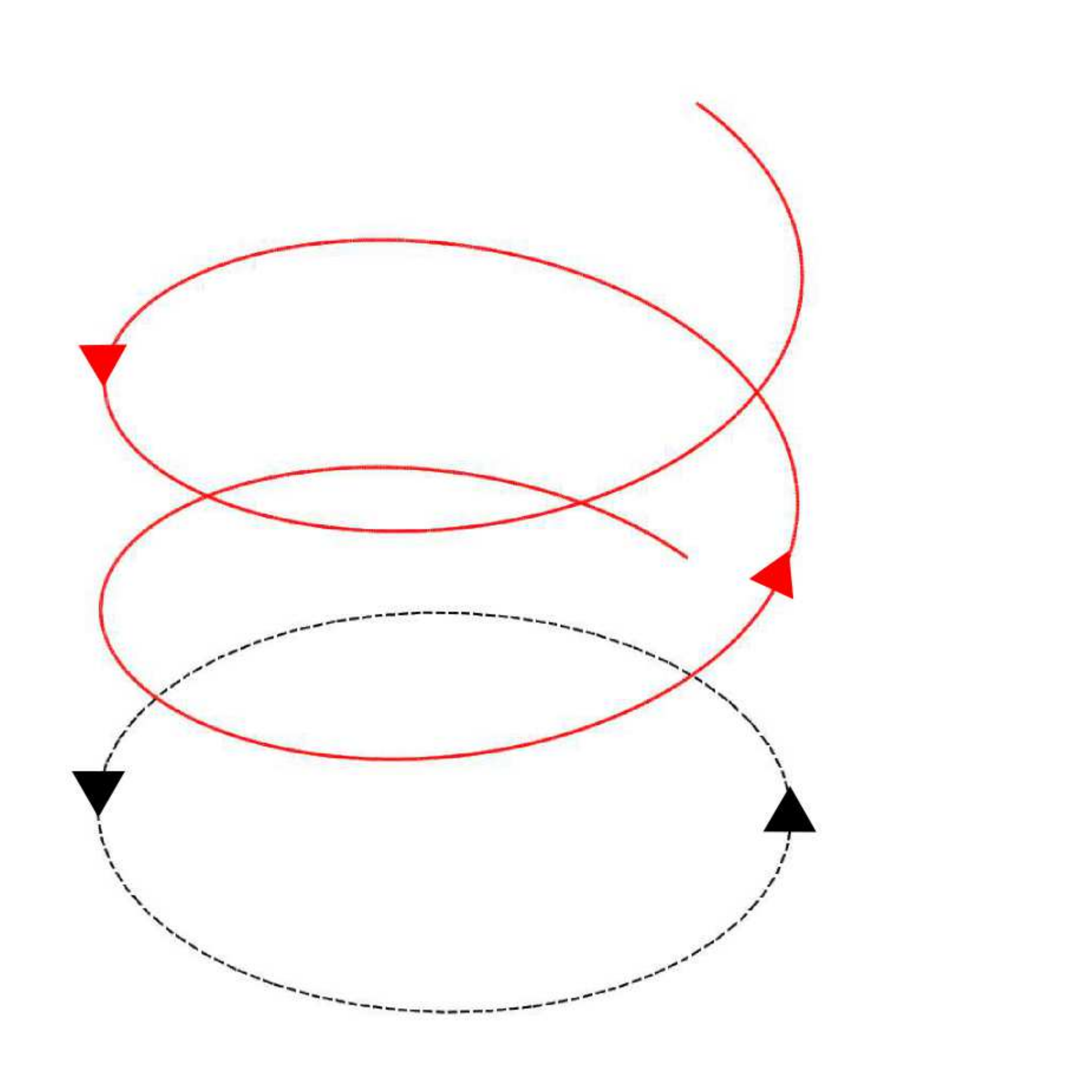}
\caption{The projection of the helical null geodesic photon 
worldline onto the spatial `circular photon orbit'. \label{F:helix}}
\end{figure}

It is important to note that this closed circular orbit is 
\textit{not} a closed null geodesic, such curves being 
highly pathological and most likely unphysical in General 
Relativity.\footnote{A closed time-like or null 
geodesic in 4-dimensional spacetime is associated with time 
travel (\cite{Carroll, Wald}, see Ref.~\cite{Lockwood} for a popular 
exposition).}  Indeed, it is not even a 
geodesic nor is it a 
null curve. Instead it is the spatial projection of an open and infinitely extended null geodesic curve in space and time as shown in Fig.~\ref{F:helix}.

The tangent vector to the null geodesic is given by
\be
\frac{dx^\mu(s)}{ds}=\left(3E,\,0,\,0,\,\frac{L}{9G^2M^2}\right)
\ee
in the spherical coordinate basis whose integral curve is the 
helical null geodesic in spacetime
\be
x^\mu(\lambda)=\left(3E\lambda, \, 0,\,0\, 
\frac{L}{9G^2M^2}\lambda\right) \,. 
\ee

A photon sphere is a sphere of radius equal to that of the 
closed circular photon orbit and coincides with a 2-sphere 
of symmetry of the spacetime. The photon sphere (a spacelike 
hypersurface) of radius $3GM$ has nothing to do 
with the black hole horizon (a null hypersurface) of radius $2GM$, 
which traps all particles.

\section{Examples of fluid flows and their associated 
Lorentzian geometries} \label{S:4}

In this section we consider various background fluid 
flows and compute the associated null curves of the auxiliary 
metric as defined in \eqref{E:E}. The purpose of this section is to use examples to highlight the distinction between the closed null curves and any kind of photon sphere. 
In an attempt to construct simple examples possessing closed 
curves uniformly stretched by a flow, we consider 
rotational symmetry. It is clear that circles concentric about 
the origin are uniformly stretched in circularly 
symmetric flows and hence constitute examples of the curves sought 
in Ref.~\cite{Haller:2013wpa} as solutions to the optimization 
problem and shown to define `coherent material vortices'.  We 
consider sequentially more complex flows: rigid body, irrotational, 
and some rotating and draining vortex flow.

\subsection{Rigid body rotation}

Consider a rigid body flow profile
\be
\phi_t:(r_0,\,\theta_0)\longrightarrow (r_0, \, \theta_0+\omega t)
\ee
where $\omega$ is a fixed angular frequency.  Then 
$\mbf{C}=\one$ coincides with the identity, $\lambda_1=\lambda_2=1$, and the 
 effective metric $\mbf{E}_\lambda$ is not of 
Lorentzian signature for any $\lambda$. Hence we must move to a more complicated flow in order to explore the construction. In a certain sense, \text{every possible curve} is null in this geometry, highlighting the total uniform and coherent (non-deforming) nature of the flow.  Such a flow is highly unphysical and does in no way invalidate the results of \cite{Haller:2013wpa}: it serves only as a first simple example which shows that the metric construction can be non-trivial.

\subsection{Irrotational vortex}

Consider instead the differential rotating flow
\be
\phi_t:(r_0,\,\theta_0)\longrightarrow \left(r_0, \, 
\theta_0+\frac{ t}{r_{ 0}}\right). \label{E:irrot}
\ee
Such a flow is irrotational and is commonly refered to as a `vortex line' flow, in this case of strength $\Gamma=2\pi$. 

\begin{figure}
\centering
\includegraphics[scale=0.4]{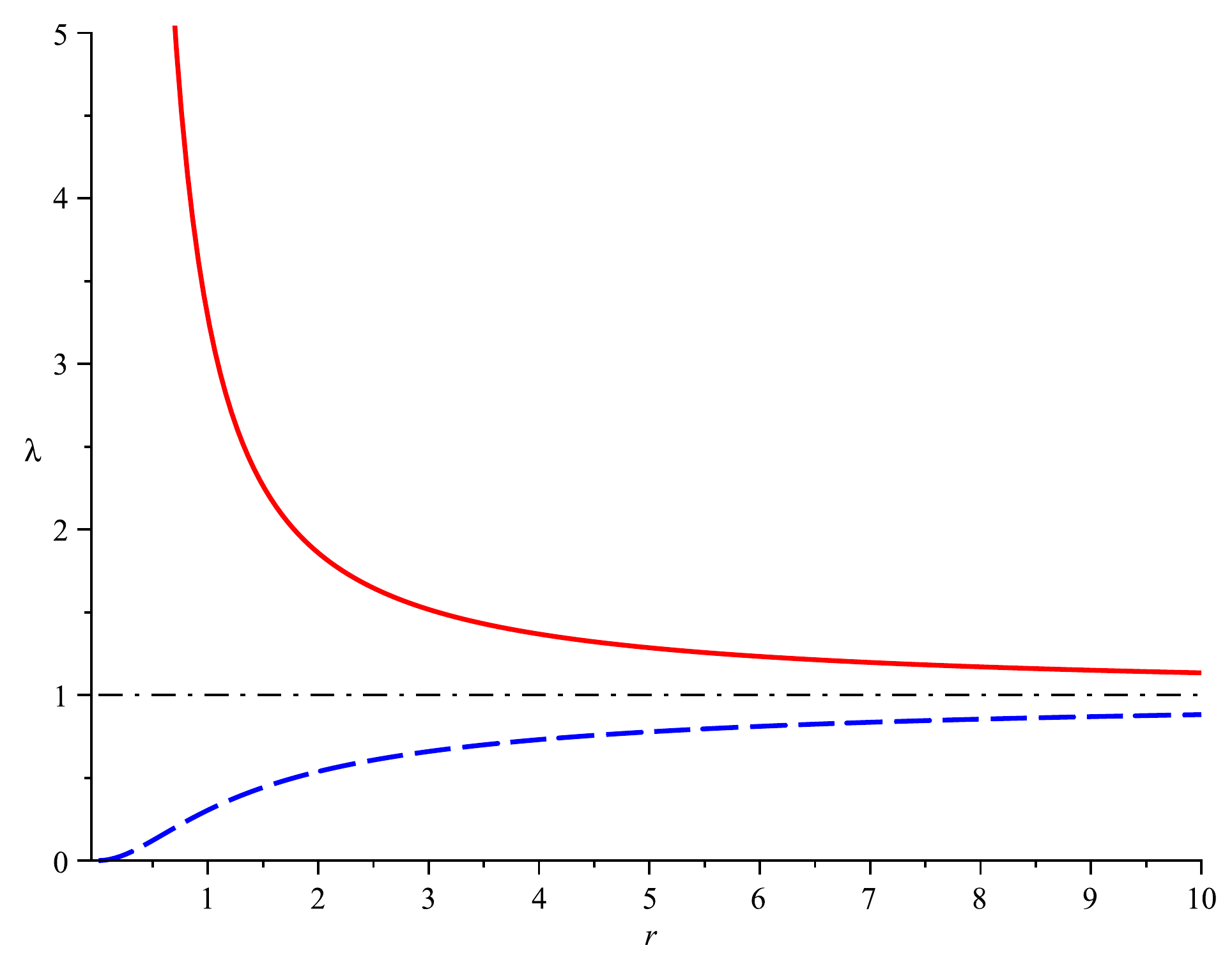}
\caption{The eigenvalues $\lambda_{\pm}$ of the matrix $\mbf{C}$ as a function of $r_0$ at finite time $t$ for the irrotational vortex flow \eqref{E:irrot}. As time proceeds the red solid curve diverges to $+\infty$ while the blue dashed curve converges to $0$ -- neither curve crosses the black dot-dashed line $\lambda=1$ at any time $t$. \label{F:values}}
\end{figure}

Then the eigenvalues of $\mbf{C}$ are easily calculable as
\be
\lambda_\pm { \left( r_0 ,  t  
\right)} =\frac{1}{{2r^2_{{ 0}}}}\left[2r^2_{{ 0}}+t^2\pm 
t\sqrt{4r^2_{{ 0}}+t^2} \right]
\ee
and we see that they are independent of the angular 
variable $\theta_0$. They satisfy, as functions of { 
$t$ and} $r_0$, the bounds
\begin{align}
&1<\lambda_+\rightarrow + \infty \quad\text{as} \quad t\rightarrow \infty \,,\\
&1>\lambda_-\rightarrow 0\quad \text{as}\quad t\rightarrow \infty \,,
\end{align}
as well as the bounds at fixed $t$
\begin{align}
&\lambda_+\rightarrow 1^+ \quad \text{as}\quad r_0\rightarrow \infty \,, \\
&\lambda_-\rightarrow 1^-\quad \text{as}\quad r_0\rightarrow \infty \,.
\end{align}
(see Fig.~\ref{F:values}). 
Hence the only value of $\lambda$ for which 
$\mbf{E}_\lambda$ is of { Lorentzian} signature in the 
entire plane is $\lambda=1$. 

\begin{figure}
\centering
\includegraphics[scale=0.4]{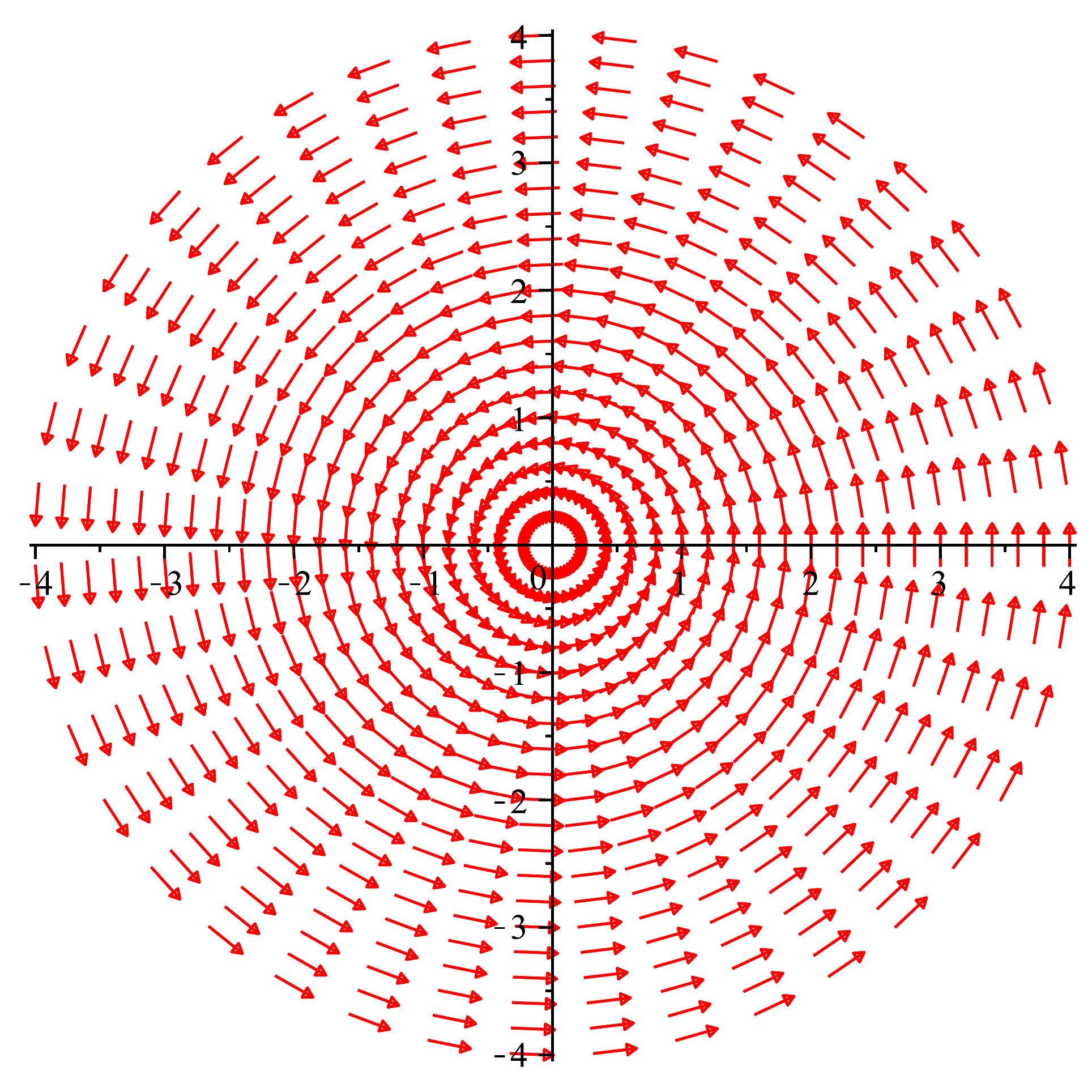}
\caption{The null vector field $\eta_+$ associated with the irrotational vortex flow \eqref{E:irrot}. This vector field is independentr of the time $t$ up to which the flow is computed. \label{F:nulls1}}
\end{figure}

\begin{figure}
\centering
\includegraphics[scale=0.4]{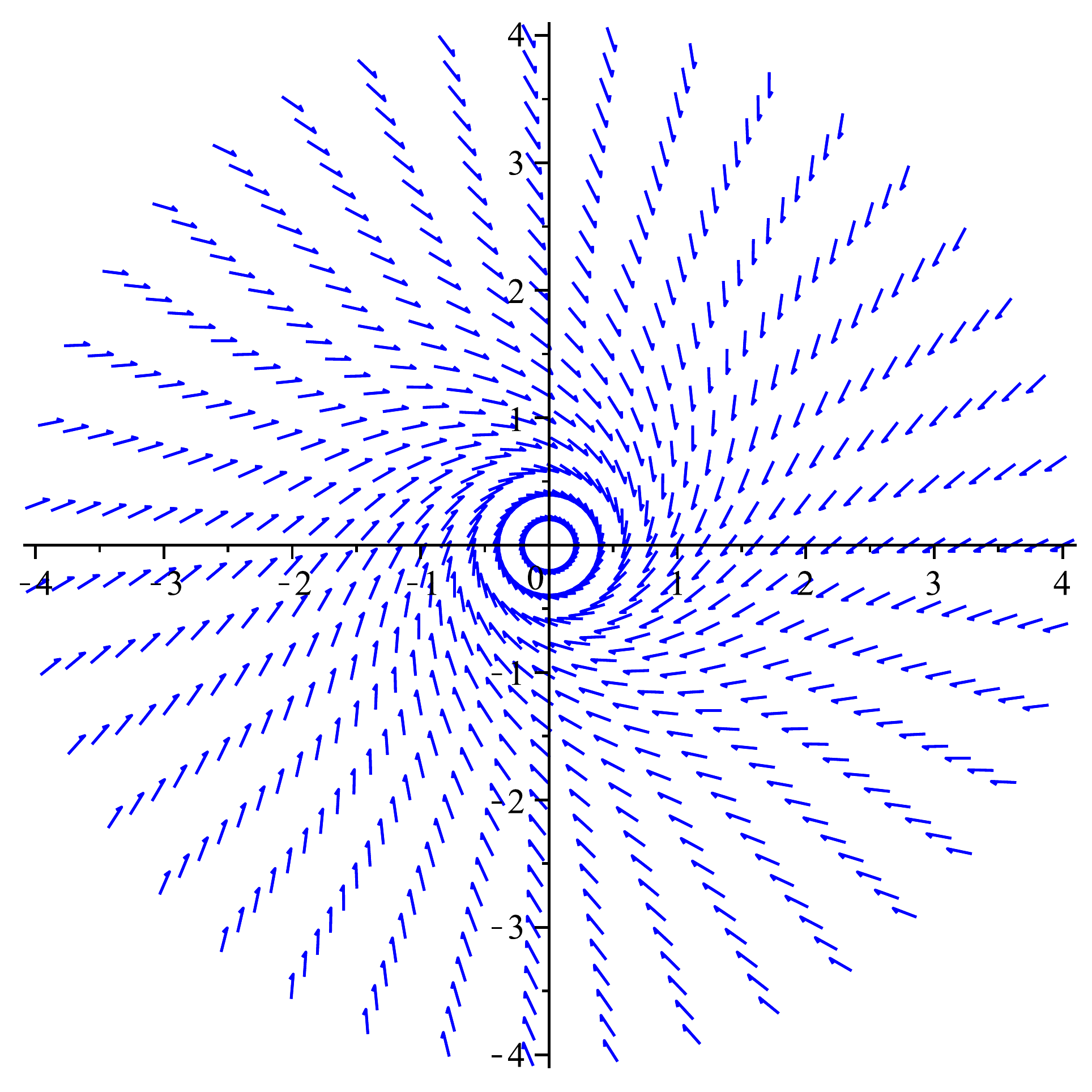}
\caption{The (normalised to unit Euclidean length) second null vector field $\eta_{-}$ \ref{E:eta-} (here at time $t=2$) associated with the irrotational flow \eqref{E:irrot}. This additional null vector field depends on the time $t$ up to which we compute the flow. At late times, or small radius it converges to $-\eta_+$, i.e. tangent to concentric circles about the origin while at early times it is purely radially pointing. \label{F:nulls2}}
\end{figure}

In the orthonormal polar basis for this choice of $\lambda$ we have
\be
E_{ab}=\frac{1}{2}\begin{pmatrix} \left(\frac{t}{r}\right)^2 & -\frac{t}{r} \vspace{2mm}\\ -\frac{t}{r} & 0\end{pmatrix}
\ee
In this case, one of the two null vector fields of 
$\mbf{E}$ is time-independent and tangent to circles 
concentric on the origin (see Fig.~\ref{F:nulls1}), in the orthonormal polar basis,
\be
\eta_+=\left(0,\,1\right)
\ee
while the other is time-dependent, 
\be
\eta_-=\left( 1,\,\frac{t}{2r}\right) \label{E:eta-}
\ee
tending to a purely angular pointing (in the $e_\theta$ direction) vector field as 
$t\rightarrow +\infty$ (see Fig.~\ref{F:nulls2}).
  
It is straightforward to show that the geodesic equation (one might find it easier to work in the non-orthonormal polar coordinate basis at this stage) becomes the pair
\begin{align}
&\frac{d^2 r(s)}{ds^2}=0 \\
&\frac{d^2\theta(s)}{ds^2}+\frac{t}{r^3}\left(\frac{dr(s)}{ds}\right)^2=0
\end{align}
from which we see that our null curves of constant $r$ are indeed null geodesics. Furthermore one can simply show that also the second set of null curves are geodesics. 

Using standard techniques one can show that there are three Killing vector fields, two of which are `spacelike' for all $t$ and $\mbf{x}_0$ and one of which is null for all $t$ and $\mbf{x}_0$.  Here `spacelike' is an arbitrary definition which we take to mean positive norm (with respect to the metric $\mbf{E}$). The null Killing vector field is purely rotational being tangent to circles about the origin. Since there is no  
1-dimensional line on which any of the  Killing vectors change norm we can conclude that there does not exist a horizon in this geometry.\footnote{Note that in lower dimensions a horizon is still defined by the change in norm of a time-like Killing vector, but is not necessarily given by a co-dimension 2 hypersurface.} 

Again, this flow is highly idealised being unbounded at the origin whereas in practice when working with real turbulent flows one would have a region $U_\lambda$ which might even be time dependent.

\subsection{Irrotational draining vortex}

One might consider also a \textit{draining} vortex type fluid flow
\be
\phi_t:(r_0,\,\theta_0)\longrightarrow 
\left(\alpha(t)\,r_0, \, \theta_0+\frac{ t}{r_{{ 0}}} 
\right) \,.
\ee
where $\alpha(t)$ is a function which parametrises the radial flow, in the hopes of unearthing some Lorentzian geometry which more closely resembles that of a black hole. Indeed, with the addition of radial flow one might hope to have some kind of ``trapped region'' inside of which all time-like vectors point towards the origin, reminiscent of a similar construction in the singularity theorems of Hawking and Penrose\cite{Hawking:1969sw}.  For example one might choose $\alpha(t)$ such that the flow describes a draining `bathtub vortex' which to a  first order approximation can be described by $\alpha(t)=\sqrt{1-At/r_0^2}$ so that $v_r(t)=-A/r(t)$ and the radial velocity is inversely proportional to the radial position as a function of time. 

The flow and geometry given here are more complex than in the previous example and, for brevity, we shall not present them in any depth here. It can be shown, however, that at fixed time $t$ the matrix $\mbf{C}$ in this case has two eigenvalues which behave analogously to the irrotational vortex case but separated now by the time-dependent `constant' $\alpha^2(t)$. In this case, and in line with our intuition, circles concentric about the origin are invariant curves and are null geodesics of the metric $\mbf{E}_{\alpha^2(t)}$: a circle of radius $r_0$ gets mapped under the flow to circles of radius $\alpha(t)\, r_0$ so that the tangent vectors squared are uniformly scaled by $\alpha^2$. This is intuitive since an experimentalist who drops ink droplets in a perfect ring into such a flow will observe a shrinking of the ring (or an expansion, depending on the character of $\alpha$) but it will be \textit{coherent} (not deforming) as time progresses. 

%
%

\section{Conclusion and outlook}

While the characterisation of fluid vortices in terms of an 
auxiliary Lorentzian metric on fixed time slices of a 
fluid flow should be of practical utility, as pointed out in 
\cite{Haller:2013wpa}, the interpretation of 
the metric and its Lorentzian geometry as being `close 
to' or analogous to
that of a black hole is lacking. We have shown that no 
event horizons exist for simple flows which possess the 
characteristic features discussed in \cite{Haller:2013wpa} 
as being `photon sphere' or `black-hole-like'. Further, we have 
clarified the 
nature of the circular photon orbit in { 
the} Schwarzschild  geometry (the prototypical black hole 
geometry containing an horizon surface which traps massive and 
massless particles)) and shown it to 
be very different from the closed null curves which hearald  the 
`coherent material vortices' discussed in 
\cite{Haller:2013wpa}. Specifically the $\lambda$-lines discussed in \cite{Haller:2013wpa} as being analogous to the photon sphere are closed null curves while the photon sphere contains closed space-like curves. 

It is interesting to point out that the use of 
Lorentzian geometry and auxiliary metrics in fluid 
dynamics in fact has a healthy and vibrant research 
community and literature associated with it known as the 
`Analogue Gravity' program \cite{lrr-2011-3} where the effective metric discussed there has a sound physical basis, being the metric which describes the real causal structure of a physical system for various kinds of physical propagating signals such as sound waves or small perturbations.  In this manner, the metrics in analogue gravity are expected to be sufficiently `physical', for example they cannot contain closed null or  closed time-like curves.  The cutting edge of that discipline is the construction and observation of analogue event horizons or analogue photon spheres in analogue systems (see for example \cite{Rousseaux:2010md,Weinfurtner:2013zfa, Philbin:2007ji}). 

In still other corners of the literature the 
introduction of an effective metric in a fictitious space 
is the key feature of the Jacobi form of the 
Maupertuis variational principle in point particle 
mechanics \cite{Goldstein}. While in principle intriguing, 
the introduction of such effective metrics in that context (see also 
\cite{others1, others2, others3, others4, others5, others6} 
for effective metrics in different contexts) 
has, thus far, been of little utility in developing theories and 
solving practical problems. It is hoped that the characterization 
of eddies by means of an effective Lorentzian geometry will break 
this spell.

\begin{acknowledgments}
We thank G. Haller and F.J. Beron-Vera for comments on a previous version of the manuscript. This work is supported by the Natural Sciences and Engineering Research Council of Canada and by Bishop's University.
\end{acknowledgments}

\providecommand{\href}[2]{#2}\begingroup\raggedright\endgroup



\end{document}